**Title: Modal analysis of heat transfer across crystalline Si and amorphous SiO$_2$ interface**

**Authors:** Kiarash Gordiz[1,*], Murali Gopal Muraleedharan[2,*], Asegun Henry[3,**]

**Affiliations:**

[1]Department of Physics, Colorado School of Mines, Golden CO, 80401, USA

[2]School of Aerospace Engineering, Georgia Institute of Technology, Atlanta, GA 30332, USA

[3]Department of Mechanical Engineering, Massachusetts Institute of Technology, Cambridge MA, 02139, USA

*These authors have equally contributed to the paper.

**Correspondence to: ase@mit.edu

**Abstract:**

We studied the anharmonic modal contributions to heat transfer at the interface of crystalline Si and amorphous SiO$_2$ using the recently proposed interface conductance modal analysis (ICMA) method. Our results show that ~74% of the thermal interface conductance (TIC) arises from the extended modes, which occupy more than ~58% of the entire population of vibrational modes in the system. More interestingly, although the population of purely localized and interfacial modes on the SiO$_2$ side is less than 6 times the population of partially extended modes, the contribution to TIC by these localized modes is more than twice that of the contribution from partially extended modes. Such an observation, once again proves the non-negligible role of localized modes to facilitate heat transfer across systems with broken symmetries, and reiterates the fact that neglecting the contribution of localized modes in any modal analysis approach is an over-simplification of the actual mechanisms of heat transfer. Correctly pinpointing the modal contributions is of vital importance, since these values are directly utilized in determining the temperature dependent TIC, which is crucial to silicon on insulator (SOI) technologies with myriad applications such as microelectronics and optoelectronics.

**Text:**

Silicon (Si) on insulator (SOI) technology is widely employed in microelectronics, where an oxide layer, typically silica ($SiO_2$), separates the top active Si layer from the bottom bulk Si handle substrate to reduce parasitic capacitance and thereby improve device performance.[1,2] Although introduction of the $SiO_2$ layer greatly enhances the electronic performance of the system by diminishing the short-channel effects,[3] it makes heat dissipation in the system more challenging. In fact, the low thermal dissipation remains the main bottleneck in achieving higher operational speeds in micro/nanoelectronics.[4,5]

The main barriers to heat dissipation in SOI devices are largely attributed to the thermal resistances (*i*) in the bulk of the $SiO_2$ layer, and (*ii*) at the interface of Si and $SiO_2$ layers (Si/$SiO_2$ interface).[6] To assess the relative significance of these two resistances, a few studies have examined the heat transfer across the Si/$SiO_2$ interface,[6-10] with only one[6] investigating the modal contributions to the thermal interface conductance (TIC). TIC is also denoted by $G$, where $Q = G\Delta T$ ($Q$, and $\Delta T$ representing the heat flux through the interface and the temperature change across the interface (i.e., Kapitza resistance[11]), respectively). Mahajan et al.[7], Lampin et al.[8] and Chen et al.[12] all employed non-equilibrium molecular dynamics (NEMD) method to obtain a temperature jump at the interface of Si/$SiO_2$ and used the above definition of $G$ to evaluate the TIC. Stanley et al.[9,10] used a modified approach to non-equilibrium heat transfer implemented in a first-principles MD simulation to capture the TIC across the Si/$SiO_2$ interface, however neither of the above groups analyzed the modal contributions to the TIC. Only Deng et al.[6] quantified the modal contributions to the heat transfer across Si/$SiO_2$ interface by employing the wave packet (WP) dynamics approach. One drawback of using WP dynamics for modal analysis is the inherently very low temperature of the system during the simulation.[13,14] Under these low temperatures, even if anharmonic interatomic potentials are used in the simulations, fully incorporating the anharmonic interactions present at the actual higher operating temperatures in the modal analysis would be challenging.[15,16] Besides the anharmonicity considerations, in the WP analysis by Deng et al.,[6] the modal contributions were ascribed to the propagating modes on the Si side (i.e., from the crystalline Si dispersion curve). Such a treatment of vibrational modes thus neglects any possible contribution to TIC by non-propagating and/or localized modes in the system. Not accounting for the role of localized modes in interfacial heat transfer[17] is an oversimplification of the problem, since localized modes are real and do exist at systems with broken symmetries, such as interfaces. For instance, using infrared electron microscopy, Krivanek et al.[18] detected and visualized the localized modes at the interface of Si/$SiO_2$. Therefore, to understand the real mechanisms of heat transfer across the Si/$SiO_2$ interface, it is of crucial importance to determine the contributions of all the actual existing modes of vibration in the system, including the non-propagating and localized interfacial modes. Such an investigation is the focus of the study presented herein.

As was discussed, the WP dynamics is unable to capture the contribution by non-propagating and/or localized modes to TIC, since this technique, as well as the other well-known ones such as acoustic mismatch model (AMM),[19,20] diffuse mismatch model (DMM),[21-23] atomistic Green's function (AGF) approach,[16,24-28] harmonic LD based approaches,[29-31] and frequency domain perfectly matched layer (FD-PML) method [32,33] is based on the phonon gas model (PGM) description[34] of heat transfer across interfaces. According to the PGM formulation,[34] the contribution by a mode of vibration to the heat transfer across the interface is proportional to its group velocity. However, as Seyf and Henry discuss in a recent study,[35] group velocity cannot be defined for the vibrational modes that are non-propagating and/or localized, which consequently makes it impossible to calculate their contribution to the interfacial heat transfer using any approach that is based on the PGM paradigm.

Recently, we introduced the interface conductance modal analysis (ICMA), a method that allows calculating the contribution by all types of modes, either propagating and/or delocalized or non-propagating and/or localized, to the interfacial heat transfer. ICMA is based on the following expression for TIC that was derived from the fluctuation-dissipation theorem [36] by Barrat *et al.* [37] and Dominguez *et al.* [38],

$$G = \frac{1}{Ak_B T^2} \int_0^\infty \langle Q(t)Q(0) \rangle dt \qquad (1)$$

where, $A$ is the cross-sectional contact area, $k_B$ is the Boltzmann constant, $T$ is the equilibrium system temperature, $\langle \cdots \rangle$ indicates the calculation of the autocorrelation function, and $Q$ represents the instantaneous exchanged energy between materials A and B. $Q$ can be calculated as,

$$Q = -\sum_{i \in A} \sum_{j \in B} \left\{ \frac{\mathbf{p}_i}{m_i} \cdot \left( \frac{-\partial H_j}{\partial \mathbf{r}_i} \right) + \frac{\mathbf{p}_j}{m_j} \cdot \left( \frac{\partial H_i}{\partial \mathbf{r}_j} \right) \right\} \qquad (2)$$

where $i$ and $j$ represent the indices for the atoms in the system, the position and momentum of the atoms in the system are denoted by $\mathbf{r}_i$ and $\mathbf{p}_i$, respectively, and the individual atom Hamiltonian is shown as $H_i$, which can be described as,

$$H_i = \frac{\mathbf{p}_i^2}{2m_i} + \Phi_i(\mathbf{r}_1, \mathbf{r}_2, \cdots, \mathbf{r}_n) \qquad (3)$$

In the above definition, $\Phi_i$ is the single atom potential energy[39,40], and $m_i$ is the mass of atom $i$. The transport picture described in Eq. (1) is based on the degree of correlation in the instantaneous interfacial heat flux, which can be captured in a molecular dynamics (MD) simulation. The key novelty in the ICMA formulation, however, is the projection of the interfacial heat flux onto the individual modes of the system.[41] This projection can be achieved by first calculating the normal mode coordinate of velocity for mode $n$ ($\dot{X}_n$) as,[42]

$$\dot{X}_n = \sum_i \frac{m_i^{1/2}}{N^{1/2}} \dot{\mathbf{x}}_i \cdot \mathbf{e}_{n,i}^* \tag{4}$$

where, $\dot{\mathbf{x}}_i$, is the velocity of atom $i$, $N$ is the total number of atoms, $*$ represents complex conjugate, and $\mathbf{e}_{n,i}$ is the eigen vector for mode $n$ describing the direction and displacement magnitude of atom $i$. The inverse operation in Eq. (4) describes the velocity of atom $i$ as the summation of individual contributions by different modes of vibration in the system as,

$$\dot{\mathbf{x}}_i = \sum_n \frac{1}{(Nm_i)^{1/2}} \mathbf{e}_{n,i} \dot{X}_n . \tag{5}$$

Replacing Eq. (5) in the definition for interfacial heat flux (Eq. (2)) results in the following definition which is the contribution by mode of vibration $n$ to the total interfacial heat flux,

$$Q_n = \frac{1}{N^{1/2}} \sum_{i \in A} \sum_{j \in B} \left\{ \left( \frac{1}{(m_i)^{1/2}} \mathbf{e}_{n,i} \dot{X}_n \right) \cdot \left( \frac{\partial H_j}{\partial \mathbf{r}_i} \right) + \left( \frac{1}{(m_j)^{1/2}} \mathbf{e}_{n,j} \dot{X}_n \right) \cdot \left( \frac{-\partial H_i}{\partial \mathbf{r}_j} \right) \right\}, \tag{6}$$

where $Q = \sum_n Q_n$. Employing the above expression for $Q_n$ in the definition of total conductance (Eq. (1)) yields the following definition for the individual modal contributions to TIC,

$$G_n = \frac{1}{Ak_B T^2} \int \langle Q_n(t) Q(0) \rangle dt , \tag{7}$$

where, $G = \sum_n G_n$. The modal contributions to TIC ($G_n$) is usually represented in the form of TIC accumulation function (see Fig. 2c). Replacing both of the total heat fluxes in Eq. (1) with their modal contributions from Eq. (6) results in the following definition, which is the individual contributions from pairs of modes of vibration equal to TIC,

$$G_{n,n'} = \frac{1}{Ak_B T^2} \int \langle Q_n(t) Q_{n'}(0) \rangle dt. \quad (8)$$

The above equation provides a quantitative description for the degree of correlation between different pairs of modes of vibration in the system, which is usually reflected on a correlation map in our calculations (see Fig. 2d). In addition, the knowledge of modal contributions by individual modes allows us to do quantum correction to get the temperature dependent TIC. Lv and Henry showed via the Green-Kubo modal analysis (GKMA) formalism [43] that results of classical MD can be extended to any temperature by correcting the heat capacity portion of the transport coefficient, leading to quantitatively correct values for modal contributions to the transport property. Their recent calculations [43,44] have given reasonably accurate results and, therefore, we apply a similar scheme here to account for quantum effects on the heat capacity of vibrational modes with frequency ($\omega$) at temperature ($T$) using the following relations,

$$G_{n,n'} = \frac{1}{Ak_B T^2} \int \langle Q_n(t) Q_{n'}(0) \rangle dt \quad (9)$$

$$G_Q(T) = \sum_\omega G_{MD}(\omega, T) \cdot \frac{C_Q(\omega, T)}{C_{MD}}, \quad (10)$$

where,

$$\frac{C_Q(\omega, T)}{C_{MD}} = \frac{\hbar \omega}{k_B} \frac{df(\omega, T)}{dT} = \frac{x^2 e^x}{(e^x - 1)^2}. \quad (11)$$

The indices $Q$ and $MD$ stand for quantum and molecular dynamics (classical), respectively. In Eq. (10) and (11), $k_B$ and $\hbar$ are Boltzmann constant and Planck's constant divided by $2\pi$, respectively, and $x = \frac{\hbar \omega}{k_B T}$. From the presented formulations, it can be seen that, contrary to all the PGM based techniques, ICMA is not based on any restrictive assumptions, thus it can be employed to study the anharmonic modal contributions of all types of vibrational modes in the system, including the localized interfacial modes, to the heat transfer across all types of interfaces, including crystalline,[45] amorphous[46] and alloyed[47] ones.

Based on the degree of localization of the vibrational energy, all the modes of vibration in an interfacial system can be classified into four groups of modes, namely <1> extended modes, <2> partially extended modes, <3> isolated modes, and <4> interfacial modes.[14] Extended modes are delocalized in the entirety of the system (Fig. 2a). The vibrations of a partially extended mode are on one side of the interface, however the vibrations only partially extend through the interface and to the other side (Fig. 2b and c). Isolated modes exist only on one side of the interface and do not include participation near the interface (Fig. 2d). Interfacial modes are localized/peaked near the interface and majorly incorporate interfacial atoms into their vibrations (Fig. 2e). We confirmed the existence of these types of vibration across the Si/SiO$_2$ interface, and typical examples of them are illustrated in Fig. 1. The percent population and the percent contribution of these modes to TIC across the Si/SiO$_2$ interface is shown in Table. 1. The nature of the contributions by these modes to TIC can be fundamentally different from the intuitions derived from the PGM paradigm. For instance, recent studies based on the ICMA technique have shown that non-propagating interfacial modes can exhibit extremely large contributions and can serve as a bridge between the two materials, by coupling to many other modes in the system.[46-48] Thus, one of the key benefits of using ICMA is that it can quantify the contribution by any of the modes that exist in an interfacial structure, which can have a variety of different types and mode characters. Therefore, in this study we used the ICMA method to not only quantify, but understand the various contributions to TIC from different vibrational modes in the Si/SiO$_2$ system.

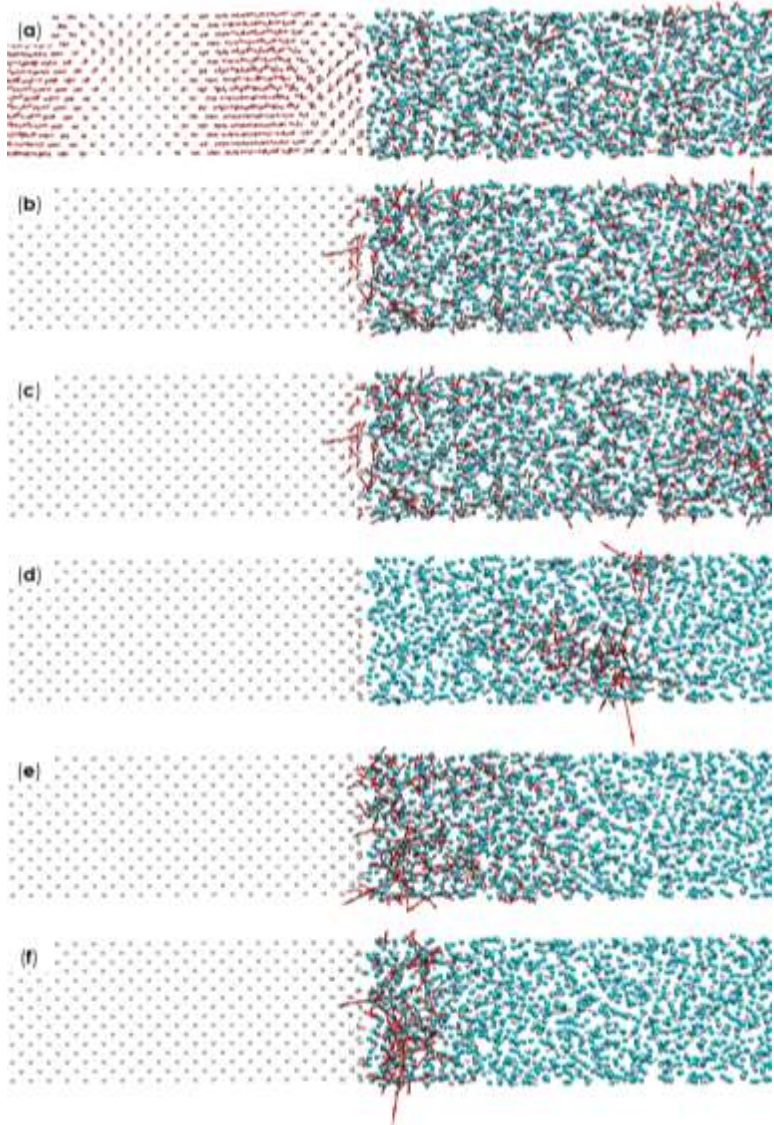

**Figure 1. Eigen vectors for different examples of the four classes of vibration present at the interface of Si/SiO$_2$.** Si and O atoms are represented by white and cyan spheres, respectively. The examples, the region they belong to in Fig. 2, and their frequencies are (**a**) extended mode in region 1 at 5.16 THz, (**b**) partially extended mode in region 2 at 23.81 THz, (**c**) partially extended mode in region 4 at 33.86 THz, (**d**) isolated mode in region 3 at 27.04 THz, (**e**) interfacial mode in region 3 at 29.85 THz, and (**f**) interfacial mode in region 5 at 32.76 THz.

**Table 1.** Number of states for the four different classes of vibration and their contribution to TIC across the Si/SiO$_2$ interface. Columns 2-4 represent the fraction of the total number of states ($\overline{DOS}$), the percentage contribution to $G$ ($\overline{G}$), and contribution to $G$ divided by fraction of total number of states (i.e., contribution to $G$ per mode) ($\overline{G}/\overline{DOS}$), respectively.

| Mode Type | $\overline{DOS}$(%) | $\overline{G}$(%) | $\overline{G}/\overline{DOS}$ |
|---|---|---|---|
| Extended | 58.14 | 74.08 | 1.27 |
| Partially extended | 35.68 | 7.97 | 0.22 |
| Isolated | 3.27 | < 0.01 | < 0.01 |
| Interfacial | 2.91 | 17.95 | 6.17 |

We used ICMA and equilibrium MD to calculate the modal contributions to TIC. Periodic boundary conditions were applied to all three Cartesian coordinates, and a time step of 0.15 fs was chosen for the MD simulations. To describe the atomic interactions, we used the Tersoff interatomic potentials.[49] We used a 3x3x30 structure (14nm length) comprising of 1664 atoms on Si side and 2160 atoms on $SiO_2$ side. We examined the effect of larger cross sections up to 5x5 unit cells and longer systems up to 50 unit cells, and neither resulted in changes larger than 5% to both the mode distributions from lattice dynamics (LD) or modal contributions to TIC calculated from MD. To generate the structure for the amorphous $SiO_2$ side, Si and O atoms are initially placed at their crystalline positions corresponding to the Cristobalite lattice. The system is then heated to a temperature above its melting point, after which it is quenched to 0 K over a 50 ns simulation time. The two sides are then brought into contact, and the entire system is annealed at 1000 K for 2 ns. This annealing/sintering process is required to ensure the correct positioning of the atoms around their equilibrium sites.[6,46] After relaxing the structure under the isobaric-isothermal ensemble (NPT) for 1 ns at zero pressure and $T=300K$ and under the canonical ensemble (NVT) for another 1 ns at $T=300K$, we simulated the structure in the microcanonical (NVE) ensemble for 10 ns during which the modal contributions to the heat flux across the interface were calculated. The heat flux contributions were saved and post processed to calculate the mode-mode heat flux correlation functions.[41] Statistical uncertainty, due to insufficient phase space averaging, has been reduced to less than 5% by considering 5 independent ensembles.[50] The MD simulations were performed using the Large Atomic/Molecular Massively Parallel Simulator (LAMMPS),[51] and the LD calculations were performed using the General Utility Lattice Program (GULP).[52]

Figure 2a shows the total and the partial density of states (DOS) belonging to each class of vibration. As expected, the low frequency modes of vibration are dominated by extended modes, and the majority of localized mode appear above ~16THz, which is the maximum frequency of vibration for the Si side. A more detailed picture for localization can also be obtained by considering the inverse participation ratio (IPR)[17,53-55] (shown in Fig. 2b), which is a direct index for localization of vibrational states in a system. A value of IPR ~ 1 corresponds to a highly localized state, and a value of IPR ~ 0 corresponds to an extended (i.e., delocalized) mode. Based on the IPR index, a partially extended mode might have an IPR only slightly larger than the IPR for an extended mode. However, following the true sense of localization, throughout this report, we will refer to a partially extended mode as a localized one, since it is localized to one section of the system, even if that section comprises half of the simulation domain. As evident from Fig. 2b, strongly localized modes are mainly distributed in three distinct frequency regions. The first region of localization is a direct effect of the interface and happens to be around the maximum frequency of the Si side (~16THz). The second (from ~25THz to ~33THz) and third (>~35THz) regions of localization are mainly due to the amorphous nature of $SiO_2$ side, where these two distinct localization regions in the bulk of amorphous $SiO_2$ correspond to locons[53] and have also been reported in previous studies.[44] In our mode classification scheme, the modes in the second and third regions of localization will be named interfacial, if they are in the proximity of the interface, and will be named isolated, if they are far from the interface. Note that In addition to the localization regions discussed above, a very small population of modes with low frequencies (~1THz) are also partially extended, indicating that localization can even interestingly happen at very low frequencies.

Figure 2c shows the modal contributions in the form of TIC accumulation function, and Fig. 2d represents the pairwise interactions/correlations of all the modes in the system (calculated form Eq. 8) in the form of the modal correlation map. To better understand the contributions by different types of vibrations, we have divided the frequency domain into five different frequency regions as shown in Fig. 2. The choice of these regions is based on the partial DOS (Fig. 2a) and the IPR (Fig. 2b). As seen in Fig. 2, region 1 lies below the maximum frequency of Si (~16 THz), where >95% of the modes are extended. These extended modes contribute >70% to TIC majorly by their elastic interactions, which is reflected on the diagonal of the correlation map (Fig. 2d). Such a large contribution to TIC by extended modes can also be seen in other structures, such as InP/InGaAs,[47] where more than 70% of TIC originates from extended modes without any need to couple to other modes in the system.

Beyond region 1, no mode can be extended, since the frequencies are larger than the maximum frequency in the bulk of Si side. Similar to the interfacial modes at the crystalline Si/Ge interface,[45,56] the interfacial modes at the Si/SiO$_2$ interface (the end of region 1) are partially extended modes with large amplitudes of vibration around the interface region. However, contrary to the crystalline Si/Ge structure, where these types of modes contribute >15% to TIC, the interfacial modes across the Si/SiO$_2$ interface at 16 THz contribute <2% to TIC. These observations substantiate that, contrary to our strong understanding of thermal conductivity,[57] thermal transport across interfaces is much more intricate,[45,58,59] which is why every interface is unique, and any generalization based on prior observations to new and unexplored interfaces might be proven to be incorrect. Regions 2 and 4 are mostly comprised of partially extended modes on the SiO$_2$ side. Despite the large population of modes in these regions (>30% of the total number of modes), their contribution to conductance is low (< 8%). On the other hand, although regions 3 and 5 do not have a large population of modes (< 5% of the total number of modes), their contribution to conductance is substantial (>15%). The main reason for such a noticeable contribution is that most of the contributing modes in regions 3 and 5 are purely localized interfacial modes (i.e., locons[17,53] on the SiO$_2$ side that are close to the interface). In fact, these interfacial modes have a very strong ability to interact with other modes of vibration to transfer their energy to the other side of the interface; this can be understood from the strong cross-correlations that these modes exhibit with other modes of vibration, which is reflected on the correlation map (Fig. 2d).

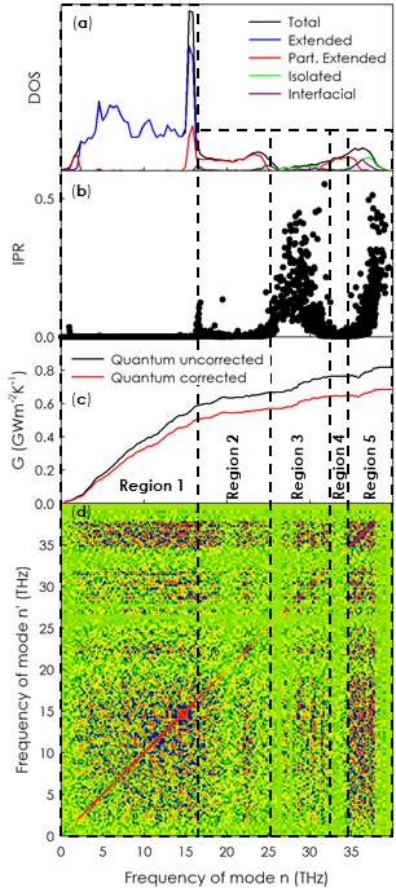

**Figure 2. DOS, IPR and modal contributions to TIC for Si/SiO$_2$ interface at T=600K, which is a typical operating temperature in SOI technology.[60]** (**a**) total DOS and DOS for different mode classifications across the interface, (**b**) IPR for the vibrational modes in the system, (**c**) quantum corrected and quantum uncorrected TIC accumulation functions, and (**d**) 2D map showing the magnitudes of the correlations/interactions on the plane of two frequency axes. The values presented on the 2D map have units of MWm$^{-2}$K$^{-1}$.

As explained previously, the quantum correction scheme can be used to correct the modal contributions calculated from the MD simulations at that specific temperature to obtain the temperature dependent TIC values, the results of which are shown in Fig. 3 alongside the temperature dependent heat capacity values.[61] Both TIC and heat capacity increase with temperature. By comparing the temperature dependent TIC and heat capacity plots, the observed increase in TIC can be attributed to the rise in heat capacity, attributed to the modes with high frequencies (mostly on $SiO_2$ side) getting excited at high temperatures. Our modal analysis depicted in Fig. 2 showed that there is a significant contribution to TIC from high frequency interfacial modes. Thus, by raising the temperature, these interfacial modes are expected to get excited and contribute to TIC. After the initial steady rise, the slope of $G(T)$ decreases at T ~500K. This is a consequence of the fact that partially extended modes around ~20-25 THz start to get excited at this temperature. However as was discussed in Fig. 2, partially extended modes in this region inherently do not contribute significantly to the TIC, thus even if they start to get excited, they only slightly change the value of TIC.

To evaluate the relative importance of thermal resistances across the $Si/SiO_2$ interface and in the bulk of $SiO_2$, we can use the value of TIC at a typical operating temperature of 600K,[60] which is ~650 MW/m$^2$-K. Considering such a value for TIC and knowing that the thermal conductivity of $SiO_2$ is ~1.2W/m-k,[62] the only way $SiO_2$ layer can have a thermal resistance smaller than the $Si/SiO_2$ interfacial resistance is by having a thickness less than 2 nm. Such a thickness may not be feasible as the typical industry-standard thickness values for the buried oxide layer are larger than 50 nm,[63] which leaves the conduction through the bulk of the $SiO_2$ layer as the main resistance to heat dissipation in SOI structures.

To confirm the accuracy of our calculations, we used three well-known thermal transport measurement techniques viz., time domain thermoreflectance (TDTR), frequency domain thermoreflectance (FDTR), and 3-omega techniques. These techniques are based on fitting an analytical heat solution to the actual response from the system. One of the unknown parameters in the analytical solution is the conductance at the interface. Our measurements, however, showed almost no sensitivity to the thermal resistance at the $Si/SiO_2$ interface, clearly indicating a high conductance at the interface, and concomitantly questioning the credibility of results obtained using these experimental approaches. The measurement procedures and efforts are fully explained in the supplementary material.

In conclusion, we studied the transfer of heat across the crystalline Si and amorphous $SiO_2$ interface and found that TIC is sufficiently large that interfacial resistance is highly unlikely to be the limiting heat dissipation factor in devices with insulating $SiO_2$ layers thicker than ~2 nm. Our modal analysis, using the ICMA method, showed that a large portion of TIC comes from the extended modes. In addition, purely localized interfacial modes also contribute noticeably to the TIC (>15%), although their population is not large (<5%). These findings further substantiate that the contribution of localized modes to thermal transport in systems with broken symmetries is crucial and should not be neglected.

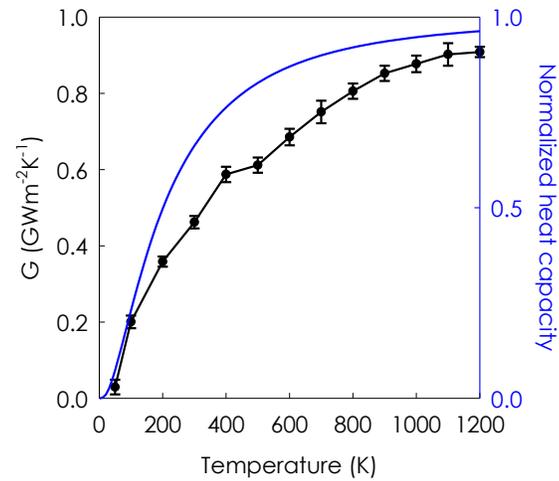

**Figure 3. Temperature dependent TIC and heat capacity in the Si/SiO2 interface structure.**

REFERENCES

1       Bruel, M. Silicon on insulator material technology. *Electron. Lett.* **31**, 1201-1202 (1995).

2       Colinge, J.-P. *Silicon-on-Insulator Technology: Materials to VLSI: Materials to Vlsi*.  (Springer Science & Business Media, 2004).

3       Veeraraghavan, S. & Fossum, J. G. Short-channel effects in SOI MOSFETs. *IEEE Trans. Electron Devices* **36**, 522-528 (1989).

4       Cahill, D. G. *et al.* Nanoscale thermal transport. II. 2003–2012. *Applied Physics Reviews* **1**, 011305 (2014).

5       VanGessel, F., Peng, J. & Chung, P. W. A review of computational phononics: the bulk, interfaces, and surfaces. *Journal of Materials Science* **53**, 5641-5683 (2018).

6       Deng, B., Chernatynskiy, A., Khafizov, M., Hurley, D. H. & Phillpot, S. R. Kapitza resistance of Si/SiO2 interface. *J. Appl. Phys.* **115**, 084910 (2014).

7       Mahajan, S. S., Subbarayan, G. & Sammakia, B. G. Estimating Kapitza Resistance Between ${\rm Si}\hbox {-}{\rm SiO} _ {2} $ Interface Using Molecular Dynamics Simulations. *IEEE Transactions on Components, Packaging and Manufacturing Technology* **1**, 1132-1139 (2011).

8       Lampin, E., Nguyen, Q.-H., Francioso, P. & Cleri, F. Thermal boundary resistance at silicon-silica interfaces by molecular dynamics simulations. *Appl. Phys. Lett.* **100**, 131906 (2012).

9       Stanley, C. M. *Heat Flow and Kapitza Resistance Across a Si/ SiO2 Interface: A First Principles Study*, (2018).

10      Stanley, C. M. & Estreicher, S. K. Heat flow across an oxide layer in Si. *physica status solidi (a)* **214**, 1700204 (2017).

11      Kapitza, P. The study of heat transfer in helium II. *J. Phys.(USSR)* **4**, 181-210 (1941).

12      Chen, J., Zhang, G. & Li, B. Thermal contact resistance across nanoscale silicon dioxide and silicon interface. *J. Appl. Phys.* **112**, 064319 (2012).

13	Schelling, P. K., Phillpot, S. R. & Keblinski, P. Phonon wave-packet dynamics at semiconductor interfaces by molecular-dynamics simulation. *Appl. Phys. Lett.* **80**, 2484-2486 (2002).

14	Gordiz, K. & Henry, A. Phonon transport at interfaces: Determining the correct modes of vibration. *J. Appl. Phys.* **119**, 015101 (2016).

15	Minnich, A. Advances in the measurement and computation of thermal phonon transport properties. *J. Phys.: Condens. Matter* **27**, 053202 (2015).

16	Mingo, N. in *Thermal nanosystems and nanomaterials*   63-94 (Springer, 2009).

17	Giri, A., Donovan, B. F. & Hopkins, P. E. Localization of vibrational modes leads to reduced thermal conductivity of amorphous heterostructures. *Physical Review Materials* **2**, 056002 (2018).

18	Krivanek, O. L. *et al.* Vibrational spectroscopy in the electron microscope. *Nature* **514**, 209 (2014).

19	Khalatnikov, I. M. *Teploobmen Mezhdu Tverdym Telom I Geliem-Ii. *Zhurnal Eksperimentalnoi I Teoreticheskoi Fiziki* **22**, 687-704 (1952).

20	Little, W. The transport of heat between dissimilar solids at low temperatures. *Can. J. Phys.* **37**, 334-349 (1959).

21	Hopkins, P. E., Norris, P. M. & Duda, J. C. Anharmonic phonon interactions at interfaces and contributions to thermal boundary conductance. *J. Heat Transfer* **133**, 062401 (2011).

22	Swartz, E. & Pohl, R. Thermal resistance at interfaces. *Appl. Phys. Lett.* **51**, 2200-2202 (1987).

23	Swartz, E. T. & Pohl, R. O. Thermal boundary resistance. *Rev. Mod. Phys.* **61**, 605 (1989).

24	Mingo, N. & Yang, L. Phonon transport in nanowires coated with an amorphous material: An atomistic Green's function approach. *Phys. Rev. B* **68**, 245406 (2003).

25	Zhang, W., Fisher, T. & Mingo, N. The atomistic Green's function method: An efficient simulation approach for nanoscale phonon transport. *Numer. Heat TR B-Fund.* **51**, 333-349 (2007).

26	Zhang, W., Mingo, N. & Fisher, T. Simulation of interfacial phonon transport in Si–Ge heterostructures using an atomistic Green's function method. *J. Heat Transfer* **129**, 483-491 (2007).

27  Tian, Z., Esfarjani, K. & Chen, G. Enhancing phonon transmission across a Si/Ge interface by atomic roughness: First-principles study with the Green's function method. *Phys. Rev. B* **86**, 235304 (2012).

28  Sadasivam, S. *et al.* Thermal transport across metal silicide-silicon interfaces: First-principles calculations and Green's function transport simulations. *Phys. Rev. B* **95**, 085310 (2017).

29  Young, D. & Maris, H. Lattice-dynamical calculation of the Kapitza resistance between fcc lattices. *Phys. Rev. B* **40**, 3685 (1989).

30  Sun, H. & Pipe, K. P. Perturbation analysis of acoustic wave scattering at rough solid-solid interfaces. *J. Appl. Phys.* **111**, 023510 (2012).

31  Zhao, H. & Freund, J. B. Phonon scattering at a rough interface between two fcc lattices. *J. Appl. Phys.* **105**, 013515 (2009).

32  Kakodkar, R. R. & Feser, J. P. A framework for solving atomistic phonon-structure scattering problems in the frequency domain using perfectly matched layer boundaries. *J. Appl. Phys.* **118**, 094301 (2015).

33  Kakodkar, R. R. & Feser, J. P. Probing the validity of the diffuse mismatch model for phonons using atomistic simulations. *arXiv preprint arXiv:1607.08572* (2016).

34  Chen, G. *Nanoscale energy transport and conversion: a parallel treatment of electrons, molecules, phonons, and photons*. (Oxford University Press, USA, 2005).

35  Seyf, H. R. & Henry, A. A method for distinguishing between propagons, diffusions, and locons. *J. Appl. Phys.* **120**, 025101 (2016).

36  Kubo, R. The fluctuation-dissipation theorem. *Rep. Prog. Phys.* **29**, 255 (1966).

37  Barrat, J.-L. & Chiaruttini, F. Kapitza resistance at the liquid—solid interface. *Mol. Phys.* **101**, 1605-1610 (2003).

38  Domingues, G., Volz, S., Joulain, K. & Greffet, J.-J. Heat transfer between two nanoparticles through near field interaction. *Phys. Rev. Lett.* **94**, 085901 (2005).

39  Hardy, R. J. Energy-flux operator for a lattice. *Phys. Rev.* **132**, 168 (1963).

40  Henry, A. S. & Chen, G. Spectral phonon transport properties of silicon based on molecular dynamics simulations and lattice dynamics. *J. Comput. Theor. Nanos.* **5**, 141-152 (2008).

41  Gordiz, K. & Henry, A. A formalism for calculating the modal contributions to thermal interface conductance. *New J. Phys.* **17**, 103002 (2015).

42  Dove, M. T. *Introduction to lattice dynamics*. Vol. 4 (Cambridge university press, 1993).

43  Lv, W. & Henry, A. Direct calculation of modal contributions to thermal conductivity via Green–Kubo modal analysis. *New J. Phys.* **18**, 013028 (2016).

44  Lv, W. & Henry, A. Non-negligible contributions to thermal conductivity from localized modes in amorphous silicon dioxide. *Scientific reports* **6** (2016).

45  Gordiz, K. & Henry, A. Phonon Transport at Crystalline Si/Ge Interfaces: The Role of Interfacial Modes of Vibration. *Scientific Reports* **6** (2016).

46  Gordiz, K. & Henry, A. Phonon transport at interfaces between different phases of silicon and germanium. *J. Appl. Phys.* **121**, 025102 (2017).

47  Gordiz, K. & Henry, A. Interface conductance modal analysis of lattice matched InGaAs/InP. *Appl. Phys. Lett.* **108**, 181606 (2016).

48  Gordiz, K. & Henry, A. Phonon Transport at Crystalline Si/Ge Interfaces: The Role of Interfacial Modes of Vibration. *Submitted to Scientific Reports* (2016).

49  Tersoff, J. Modeling solid-state chemistry: Interatomic potentials for multicomponent systems. *Phys. Rev. B* **39**, 5566 (1989).

50  Gordiz, K., Singh, D. J. & Henry, A. Ensemble averaging vs. time averaging in molecular dynamics simulations of thermal conductivity. *J. Appl. Phys.* **117**, 045104 (2015).

51  Plimpton, S. Fast parallel algorithms for short-range molecular dynamics. *J. Comput. Phys.* **117**, 1-19 (1995).

52  Gale, J. D. GULP: A computer program for the symmetry-adapted simulation of solids. *J. Chem. Soc., Faraday Trans.* **93**, 629-637 (1997).

53  Allen, P. B., Feldman, J. L., Fabian, J. & Wooten, F. Diffusons, locons and propagons: Character of atomie yibrations in amorphous Si. *Philos. Mag. B* **79**, 1715-1731 (1999).

54  Allen, P. B. & Feldman, J. L. Thermal conductivity of disordered harmonic solids. *Phys. Rev. B* **48**, 12581 (1993).

55  Larkin, J. M. & McGaughey, A. J. Predicting alloy vibrational mode properties using lattice dynamics calculations, molecular dynamics simulations, and the virtual crystal approximation. *J. Appl. Phys.* **114**, 023507 (2013).

56  Zhou, Y. & Hu, M. Full quantification of frequency-dependent interfacial thermal conductance contributed by two-and three-phonon scattering processes from nonequilibrium molecular dynamics simulations. *Phys. Rev. B* **95**, 115313 (2017).

57  Toberer, E. S., Baranowski, L. L. & Dames, C. Advances in thermal conductivity. *Annual Review of Materials Research* **42**, 179-209 (2012).

58  Giri, A. & Hopkins, P. E. Role of interfacial mode coupling of optical phonons on thermal boundary conductance. *Scientific reports* **7**, 11011 (2017).

59  Giri, A. *et al.* Small-mass atomic defects enhance vibrational thermal transport at disordered interfaces with ultrahigh thermal boundary conductance. *arXiv preprint arXiv:1710.09440* (2017).

60  Kappert, H., Kordas, N., Dreiner, S., Paschen, U. & Kokozinski, R. in *Circuits and Systems (ISCAS), 2015 IEEE International Symposium on.* 1162-1165 (IEEE).

61  Kittel, C. & McEuen, P. *Introduction to solid state physics*. Vol. 8 (Wiley New York, 1986).

62  Regner, K. T. *et al.* Broadband phonon mean free path contributions to thermal conductivity measured using frequency domain thermoreflectance. *Nature communications* **4**, 1640 (2013).

63  Celler, G. K. SOI Technology Driving the 21st Century Ubiquitous Electronics. *ECS Transactions* **19**, 3-14 (2009).